\documentclass[journal]{IEEEtran}
\usepackage{amsmath,amsfonts,amssymb}
\usepackage{graphicx}
\usepackage{algorithm}
\usepackage{algorithmic}
\usepackage{hyperref}
\usepackage{multirow}
\usepackage{caption}
\usepackage{subcaption}

\begin{document}

\title{BERT4MIMO: A Foundation Model using BERT Architecture for Massive MIMO Channel State Information Prediction}

\author{\IEEEauthorblockN{Ferhat Ozgur Catak\(^1\), Murat Kuzlu\(^2\), Umit Cali\(^3\)}\\
\IEEEauthorblockA{
\textit{\(^1\)Department of Electrical Engineering and Computer Science, University of Stavanger}, Rogaland, Norway\\
\textit{\(^2\)Department of Engineering Technology, Old Dominion University}, Norfolk, VA, USA\\
\textit{\(^3\)University of York, School of Physics, Engineering and Technology, Heslington}, York YO10 5DD, United Kingdom \\
f.ozgur.catak@uis.no, mkuzlu@odu.edu, umit.cali@york.ac.uk}}

\maketitle

\begin{abstract}
Massive MIMO (Multiple-Input Multiple-Output) is an advanced wireless communication technology, using a large number of antennas to improve the overall performance of the communication system in terms of capacity, spectral, and energy efficiency. The performance of MIMO systems is highly dependent on the quality of channel state information (CSI). Predicting CSI is, therefore, essential for improving communication system performance, particularly in MIMO systems, since it represents key characteristics of a wireless channel, including propagation, fading, scattering, and path loss.
This study proposes a foundation model inspired by BERT, called BERT4MIMO, which is specifically designed to process high-dimensional CSI data from massive MIMO systems. BERT4MIMO offers superior performance in reconstructing CSI under varying mobility scenarios and channel conditions through deep learning and attention mechanisms. The experimental results demonstrate the effectiveness of BERT4MIMO in a variety of wireless environments.

\end{abstract}


\section{Introduction}
In recent decades, wireless communication systems, particularly next-generation networks (known as NextG or 5G and beyond) have been significantly improved along with the support of industry and academia \cite{chen20235g, mahmood2021industrial, zeb2022industry}. NextG networks are the most promising technologies in this cyber era, aiming to connect billions of devices and systems on the same platform at high data rates and low latency to support new applications such as digital twins, extended reality, collaborative robots, smart grid 2.0, industry 5.0, autonomous vehicles, and intelligent healthcare services \cite{porambage2021roadmap}. With advanced communication, computing, and Artificial Intelligence / Machine Learning (AI/ML) technologies, NextG networks can successfully support these applications. AI/ML among these technologies has a unique contribution to enabling advanced applications \cite{10056645}, such as MIMO (Multiple-Input Multiple-Output) \cite{10520592}, beamforming \cite{brilhante2023literature}, spectrum sensing \cite{shi2022federated}, channel estimation \cite{catak2023cryptographic}, network slicing \cite{wang2023artificial}, Automatic Modulation Recognition (AMR) \cite{tang2024automatic}, and Intelligent Reflecting Surfaces (IRS) \cite{catak2022security} in NextG networks. 

AI-based solutions have also changed dramatically by introducing generative AI (GenAI) models, particularly Generative Pre-trained Transformer 3 (GPT-3) by OpenAI in 2020, the groundbreaking in the AI world at that time \cite{aydin2022research}. Following this point, researchers have started to explore foundation models that can serve as a base layer for various applications. In the literature, there are many studies on GenAI-based solutions that support applications in healthcare \cite{zhang2023generative, kuzlu2023rise}, finance \cite{lee2024comprehensive}, education \cite{alasadi2023generative, chere2024integration}, intelligent transportation \cite{yan2023survey, sun2024generative}, e-commerce \cite{chang2024human}, and smart manufacturing \cite{ghobakhloo2024generative}, while there are limited studies in telecommunications and NextG networks. The authors in \cite{wang2024generative} present the GenAI agent framework for NextG MIMO design with the use of a large language model (LLM) and retrieval augmented generation (RAG). They also review the development, fundamentals, and challenges of the NextG MIMO design along with how the GenAI agent facilitates the NextG MIMO applications. The survey  \cite{van2024generative} provides a comprehensive overview of the physical layer applications in communication systems using GenAI-based methods, along with challenges and future research directions to enhance them. These applications include ranging from traditional (signal classification, channel estimation, and equalization) to advanced topics (beamforming, IRS, and joint source-channel coding). It also emphasizes the unique contribution of the use of GenAI methods in NextG networks due to its capabilities to extract, transform, and enhance complex data features. The authors in \cite{alikhani2024large} propose the first foundation model for wireless channels, called the Large Wireless Model (LWM), developed for wireless communication and sensing systems. The proposed LWM is built on a transformer-based architecture, pre-trained in a foundation model on large-scale wireless channel datasets. The findings indicate that LWM can generate real-time embeddings for raw channels, effectively extracting detailed and complex patterns from the input, and adapting to diverse tasks with limited datasets. This can be considered a breakpoint as it proposes the first foundation model for wireless communication systems and future research directions in this field.

Transformer models serving as the backbone of many GenAI systems, particularly BERT (Bidirectional Encoder Representations from Transformers), have achieved remarkable success in various domains, including natural language processing (NLP) and time-series forecasting \cite{kar2023unravelling}. This paper follows this GenAI trend and focuses on developing a foundation model inspired by BERT, called BERT4MIMO, for massive MIMO channel state prediction in NextG networks.  


The main contributions of this paper are (1) A novel dataset generation using MATLAB to simulate CSI matrices for various scenarios, including stationary, high-speed mobility, and urban macrocell environments, (2) The design and implementation of a foundation model, called BERT4MIMO, utilizing a transformer-based architecture along with feature embeddings to enhance CSI reconstruction, and (3) Experimental evaluation of the robustness of BERT4MIMO under various conditions, including scenarios with Doppler shifts and different masking ratios.


\section{Preliminaries}

\subsection{Massive MIMO Channel State Information}

Massive multiple-input multiple-output (MIMO) system is one of the promising advanced technologies in wireless communication systems, using a large number of antennas at the base station (BS) to enhance capacity, spectral efficiency, reliability, and energy efficiency. The performance of massive MIMO is heavily based on accurate Channel State Information (CSI), which captures the channel characteristics, including path gains, scattering, and propagation delays \cite{9566598, 8794743}.

In a typical Massive MIMO system, the CSI information is formalized as matrix $\mathbf{H} \in \mathbb{C}^{N_s \times N_t \times N_r}$ to describe the time-varying channel characteristics between the base station (BS) and the user equipment (UE). The matrix is three-dimensional.
where:
\begin{itemize}
    \item $N_s$ is the number of subcarriers used in the frequency domain representation of the channel,
    \item $N_t$ is the number of transmit antennas at the BS,
    \item $N_r$ is the number of receive antennas at the UE.
\end{itemize}
In the CSI matrix, $h_{s,t,r}$ represents the complex channel gain from the $t_{th}$ transmit antenna to the $r_{th}$ receive antenna on the $s_{th}$ subcarrier for each entry:

\begin{equation}
h_{s,t,r} = g_{s,t,r} e^{j \phi_{s,t,r}}
\end{equation}

where:
\begin{itemize}
    \item \( g_{s,t,r} \) is the magnitude of the channel gain,
    \item \( \phi_{s,t,r} \) is the phase shift resulting from the propagation path.
\end{itemize}

The CSI matrix $\mathbf{H}$ can be split into several components due to model different propagation effects, which are represented as follows:
\begin{equation}
\mathbf{H} = \mathbf{H}_{\text{LOS}} + \mathbf{H}_{\text{NLOS}} + \mathbf{H}_{\text{Noise}}
\end{equation}
where:
\begin{itemize}
    \item $\mathbf{H}_{\text{LOS}}$ represents the line-of-sight (LOS) component, which models the direct path between the transmitter and the receiver,
    \item $\mathbf{H}_{\text{NLOS}}$ is the non-line-of-sight (NLOS) component, capturing the reflected, scattered, and diffracted paths,
    \item $\mathbf{H}_{\text{Noise}}$ denotes the additive white Gaussian noise (AWGN) matrix.
\end{itemize}

\subsection{Physical Layer Channel Models}

A channel is typically modeled using a combination of deterministic and statistical models in massive MIMO \cite{ali2009deterministic}. Defining models describe known propagation characteristics, such as LOS, while statistical models describe randomness parameters due to environmental factors, such as multipath propagation, scattering, mobility, and interference. The general form of a statistical MIMO channel model can be given as follows:

\begin{equation}
\mathbf{H} = \mathbf{A} \mathbf{B} + \mathbf{N}
\end{equation}
where:
\begin{itemize}
    \item $\mathbf{A}$ is the matrix modeling the spatial characteristics of the channel (i.e., path loss, angles of arrival/departure),
    \item $\mathbf{B}$ is the matrix representing the multipath components (i.e., reflection, scattering, diffraction),
    \item $\mathbf{N}$ is the noise matrix modeling as Gaussian noise.
\end{itemize}


The path loss for the channel between the BS and the UE is a function of the distance between the antennas ($d$), and the frequency of operation ($f_c$). The path loss $\alpha$ can be expressed with the distance and the frequency as follows:
\begin{equation}
    \alpha(d, f_c) = \frac{1}{d^\beta} \left(\frac{f_c}{f_0}\right)^\gamma
\end{equation}
where:
\begin{itemize}
    \item $\beta$ is the path loss exponent (typically between 2 and 4 depending on the environment),
    \item $f_0$ is a reference frequency,
    \item $\gamma$ is the frequency-dependent scaling factor.
\end{itemize}

Doppler shift, $\Delta f$, presents frequency shifts in the received signal for mobility cases like vehicles, and can be given as follows:
\begin{equation}  
\Delta f = \frac{v f_c}{c}
\end{equation} 
where:
\begin{itemize}
    \item $v$ is the speed of the UE,
    \item $f_c$ is the carrier frequency,
    \item $c$ is the speed of light.
\end{itemize}

In addition, this shift results in a frequency offset in the CSI, affecting the coherence time and bandwidth of the system.

\subsection{CSI Matrix Preprocessing}
For AI/ML systems, the raw CSI data needs to be preprocessed in a format suitable for model training and testing. This typically involves normalizing the real and imaginary components of the CSI data, followed by flattening and padding the matrices.

A CSI matrix was defined earlier as $\mathbf{H} \in \mathbb{C}^{N_s \times N_t \times N_r}$. First, it is needed to extract its real and imaginary parts for each entry, and then to normalize across each subcarrier:
\begin{equation} 
\mathbf{H}_{\text{Real}} = \text{Re}(\mathbf{H}), \quad \mathbf{H}_{\text{Imag}} = \text{Im}(\mathbf{H})
\end{equation} 
\begin{equation} 
\mathbf{H}_{\text{Real}}^{\text{norm}} = \frac{\mathbf{H}_{\text{Real}} - \mu_{\text{Real}}}{\sigma_{\text{Real}}}, \quad \mathbf{H}_{\text{Imag}}^{\text{norm}} = \frac{\mathbf{H}_{\text{Imag}} - \mu_{\text{Imag}}}{\sigma_{\text{Imag}}}
\end{equation} 

where $\mu_{\text{Real}}, \mu_{\text{Imag}}$ and $\sigma_{\text{Real}}, \sigma_{\text{Imag}}$ are the mean and standard deviation of the real and imaginary parts of $\mathbf{H}$, respectively.

The real and imaginary components are then stacked together to form a combined tensor, and then reshaped into a 2D tensor $\mathbf{X} \in \mathbb{R}^{N_s \times d}$, where $d = 2 N_t N_r$ is the dimension of the feature, including both the real and imaginary parts.
\begin{equation}
\mathbf{H}_{\text{Combined}} = \text{concat}(\mathbf{H}_{\text{Real}}^{\text{norm}}, \mathbf{H}_{\text{Imag}}^{\text{norm}})
\end{equation}


\subsection{Masking}
Random masking is applied to the CSI data to simulate realistic data loss in wireless communications \cite{zhao2024finding}. To achieve that, a mask matrix $\mathbf{M} \in \{0, 1\}^{N_s \times d}$ is generated, where each element $m_{s,i}$ is independently sampled from a Bernoulli distribution with a probability $p$ as follows:
\begin{equation}
m_{s,i} \sim \text{Bernoulli}(p)
\end{equation}

The masked CSI data is then obtained by element-wise multiplication with the mask:
\begin{equation}
\mathbf{X}_{\text{Masked}} = \mathbf{X} \odot \mathbf{M}
\end{equation}
This random masking simulates the scenario where parts of the CSI are unavailable due to noise, interference, or limited measurements.

\section{Dataset Details and Simulation Scenarios}

\subsection{Dataset Description}
In this study, the synthesized dataset is generated to simulate a set of scenarios in massive MIMO systems utilizing the MATLAB 5G Toolbox, employing the TDL (Tapped Delay Line) channel models for different delay profiles\footnote{https://www.mathworks.com/products/5g.html}. These scenarios reflect the variability and challenges encountered in practical deployments, including stationary users, high-speed mobility, and urban macrocell environments. Detailed specifications for the data set are provided here.

CSI matrices defined as $\mathbf{H}_{\text{c,u,s}} \in \mathbb{C}^{N_s \times N_t \times N_r}$ form the dataset, which is indexed by three identifiers:
\begin{equation}
    \mathcal{D} = \{\mathbf{H}_{c,u,s} : c \in [1, C], u \in [1, U], s \in [1, S]\}
\end{equation}

where:
\begin{itemize}
    \item $C$, $U$, and $S$ denote the total number of cells, UEs per cell, and scenarios, respectively,
    \item $N_s$: Number of subcarriers, representing the frequency domain,
    \item $N_t$: Number of transmit antennas at the base station (BS),
    \item $N_r$: Number of receive antennas at the user equipment (UE).
\end{itemize}


In MATLAB 5G Toolbox, the utilized parameters are set as follows:
\begin{itemize}
    \item $C = 10$: Number of cells (base stations).
    \item $U = 200$: Number of UEs per cell.
    \item $S = 3$: Number of scenarios (Stationary, High-Speed, Urban Macro). They will be explained. 
    \item $N_s = 64$: Number of subcarriers.
    \item $N_t = 64$: Number of transmit antennas at the BS.
    \item $N_r = 4$: Number of receive antennas at the UE.
\end{itemize}

\subsection{Dataset Generation}
For each set of the training dataset, a new channel characteristic is generated based on various channel parameters, such as delay profiles (TDL-A, TDL-C, TDL-D), delay spreads (100-500 nanosecond), Doppler shifts, and Signal-to-noise ratio (SNR) changes between 0 and 100 dB. The noise signal is generated with an SNR of 100 dB.  For each scenario, the CSI is computed for $N_s$ subcarriers, $N_t$ transmit antennas at the BS, and $N_r$ receive antennas at the UE as follows:
\begin{equation}
    \mathbf{H}_{s,t,r} = \sum_{p=1}^{P} \alpha_p \mathbf{a}_t(\theta_p) \mathbf{a}_r^\dagger(\phi_p) e^{-j2\pi f_s \tau_p}
\end{equation}
where:
\begin{itemize}
    \item $P$: Number of multipath components.
    \item $\alpha_p$: Complex gain of the $p$-th path.
    \item $\mathbf{a}_t(\theta_p)$: Transmit antenna array response for angle $\theta_p$.
    \item $\mathbf{a}_r^\dagger(\phi_p)$: Conjugate transpose of the receive antenna array response for angle $\phi_p$.
    \item $\tau_p$: Delay of the $p$-th path.
    \item $f_s$: Subcarrier frequency.
\end{itemize}

In addition, noise is added to each computed CSI to simulate real-world environments using the SNR.
\begin{equation}
    \mathbf{H}_{\text{Noisy}} = \mathbf{H} + \mathbf{N}
\end{equation}

\subsection{Dataset Preprocessing}
The following preprocessing steps are applied to the generated dataset before feeding it into the BERT4MIMO model:
\begin{itemize}
    \item \textbf{Normalization:} Each CSI matrix $\mathbf{H}$ is normalized across the dataset:
    \begin{equation}
        \mathbf{H}_{\text{Norm}} = \frac{\mathbf{H} - \mu_{\mathbf{H}}}{\sigma_{\mathbf{H}}}
    \end{equation}
    \item \textbf{Masking:} To simulate the prediction task, a random mask $\mathbf{M}$ is applied to the input:
    \begin{equation}
        \mathbf{X}_{\text{Masked}} = \mathbf{X} \odot \mathbf{M}
    \end{equation}
    where $\odot$ represents element-wise multiplication.
\end{itemize}
These steps ensure that the data is robustly prepared for the transformer-based architecture.

\subsection{Scenarios and Their Characteristics}
The dataset includes three distinct scenarios, each simulating a unique environment with specific delay profiles, Doppler shifts, and noise characteristics:
\begin{itemize}
    \item \textbf{Stationary UE:} This scenario models users in static environments, such as indoor deployments. The delay profile is modeled using TDL-A, with a short delay spread $\tau_{\text{Stationary}} = 100$ ns and no Doppler shift:
    \begin{equation}
        \Delta f_{\text{Stationary}} = 0.
    \end{equation}

    \item \textbf{High-Speed UE:} This scenario simulates users in vehicles moving at high speeds. The delay profile follows TDL-C, with a moderate delay spread $\tau_{\text{High-Speed}} = 300$ ns and a significant Doppler shift in Hz:
    \begin{equation}
        \Delta f_{\text{High-Speed}} = \frac{v \cdot f_c}{c},
    \end{equation}
    where $v$ is the UE speed (120 km/h), $f_c$ is the carrier frequency (3.5 GHz), and $c$ is the speed of light.

    \item \textbf{Urban Macrocell:} This scenario captures the characteristics of large-scale urban deployments with significant multipath effects. The delay profile is modeled using TDL-D, with a long delay spread $\tau_{\text{Urban Macro}} = 500$ ns and no Doppler shift:
    \begin{equation}
        \Delta f_{\text{Urban Macro}} = 0
    \end{equation}
\end{itemize}

The complete dataset consists of $C \cdot U \cdot S = 6,000$ CSI matrices, distributed across the three scenarios. The breakdown is provided in Table~\ref{tab:dataset_distribution}.

\begin{table}[ht]
    \caption{Dataset distribution for each scenario}
    \centering
    \begin{tabular}{|c|c|c|}
        \hline
        \textbf{Scenario} & \textbf{Number of CSI Matrices} & \textbf{Percentage} \\
        \hline
        Stationary UE     & 2,000                          & 33.33\%             \\
        High-Speed UE     & 2,000                          & 33.33\%             \\
        Urban Macrocell   & 2,000                          & 33.33\%             \\
        \hline
        \textbf{Total}    & \textbf{6,000}                 & \textbf{100\%}      \\
        \hline
    \end{tabular}
    \label{tab:dataset_distribution}
\end{table}

\section{BERT4MIMO Architecture}

The proposed \textit{BERT4MIMO} model utilizes a transformer-based architecture to process high-dimensional Channel State Information (CSI) data. It integrates domain-specific adaptations, such as temporal embeddings and feature-specific encodings, to ensure efficient learning and accurate CSI reconstruction. The source code regarding BERT4MIMO architecture is released on GitHub for scientific use\footnote{https://github.com/ocatak/BERT4MIMO-AI4Wireless}, and the following subsections describe the architecture in detail.

\begin{figure}[h]
    \centering
    \includegraphics[width=\linewidth]{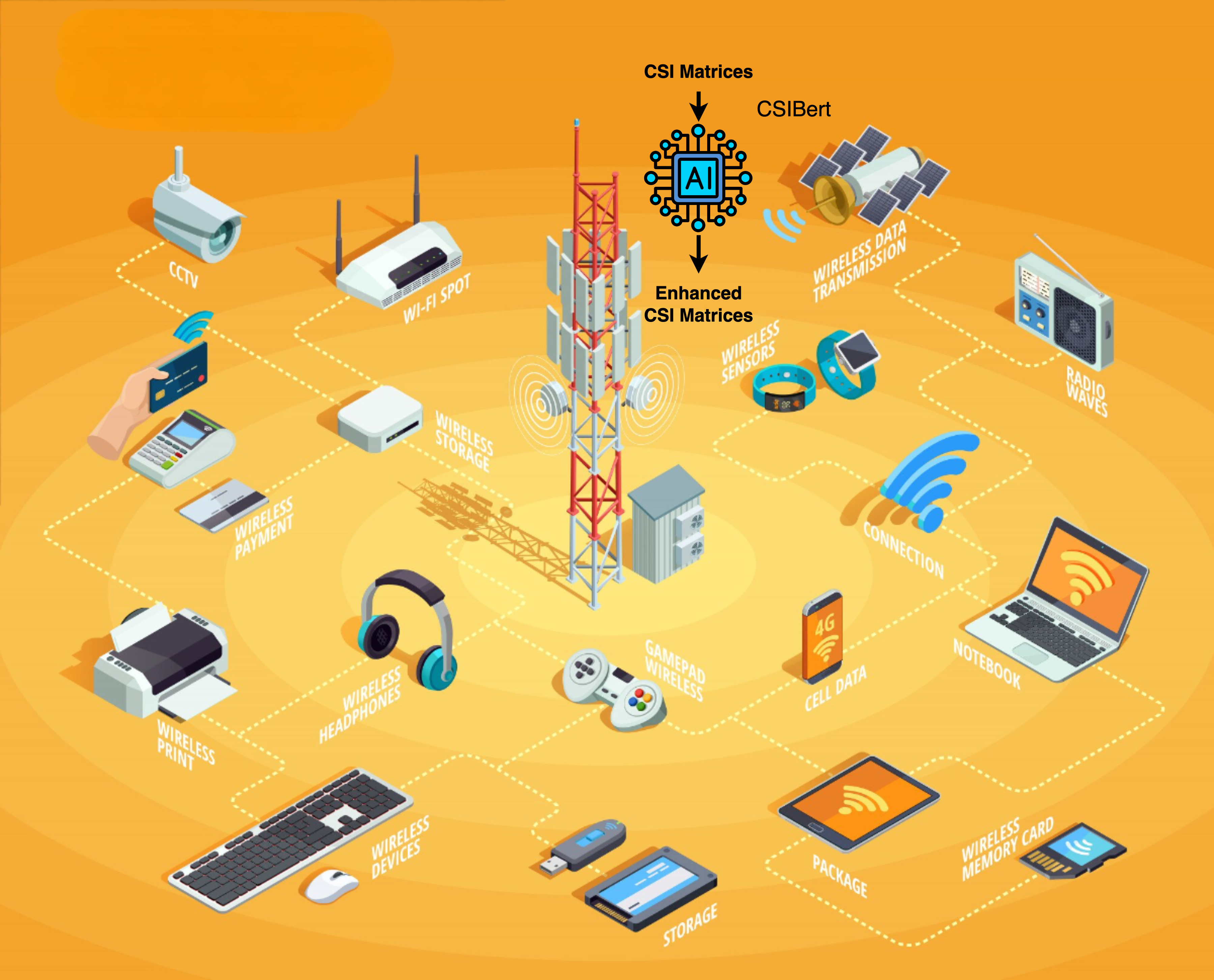}
    \caption{An illustrative overview of BERT4MIMO's role in modern wireless communication systems. 
    }
    \label{fig:overview-BERT4MIMO}
\end{figure}

Figure~\ref{fig:overview-BERT4MIMO} illustrates the high-level workflow of BERT4MIMO within a wireless communication ecosystem. At the center is the base station (BS) equipped with massive MIMO antennas, which receive CSI matrices from various connected wireless devices, such as mobile phones, IoT devices, and sensors. These CSI matrices capture the characteristics of the wireless channel, including propagation, fading, and scattering.

Using BERT4MIMO, the raw CSI data is processed to reconstruct and enhance the CSI matrices with real and imaginary components. The enhanced CSI matrices are then utilized to improve communication reliability, spectral efficiency, and energy optimization across various application areas, including smart cities, intelligent transportation, and industrial IoT. 

\begin{figure}[!htbp]
    \centering
    \includegraphics[width=0.85\linewidth]{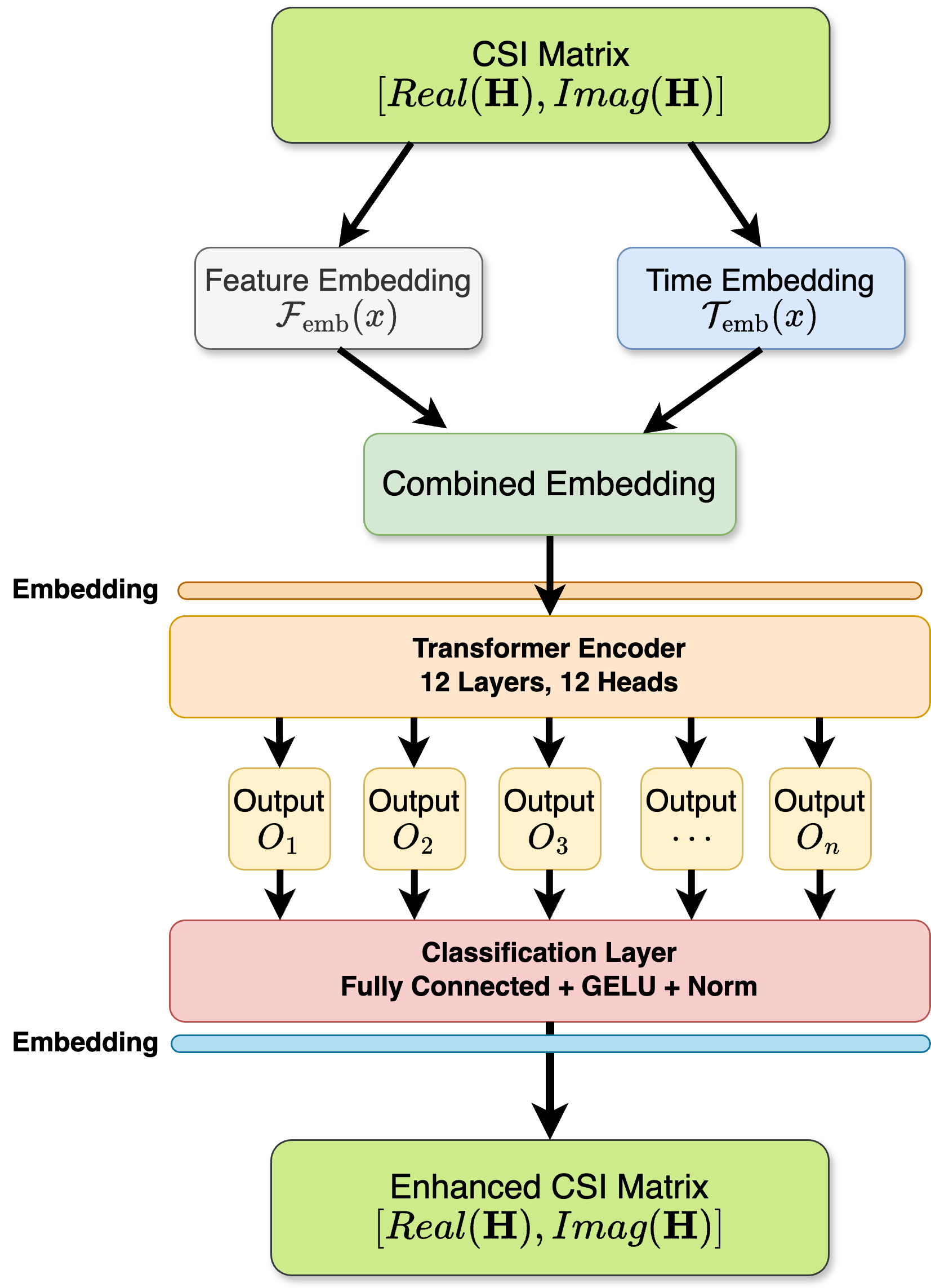} 
    \caption{The architecture of BERT4MIMO
    }
    \label{fig:BERT4MIMO_architecture}
\end{figure}

The architecture of BERT4MIMO, illustrated in Figure \ref{fig:BERT4MIMO_architecture}, begins with two embedding layers: the \textit{Feature Embedding} layer $\mathcal{F}_{\text{emb}}(x)$, which encodes spatial and frequency dimensions, and the \textit{Time Embedding} layer $\mathcal{T}_{\text{emb}}(x)$, which encodes temporal information. These embeddings are combined to form a unified representation, referred to as the \textit{Combined Embedding}, which captures both temporal and spatial features. This combined embedding is processed by a Transformer encoder, consisting of 12 layers with 12 attention heads per layer. The encoder captures dependencies between different subcarriers and antennas across both time and frequency domains.

Each output $\mathbf{O}_i$ from the Transformer encoder corresponds to an intermediate representation of the CSI data, which is then passed through the classification layer. The classification layer, comprising a fully connected network with GELU activation and normalization, reconstructs the Enhanced CSI Matrix $\hat{\mathbf{H}} = [\text{Re}(\mathbf{H}), \text{Im}(\mathbf{H})]$. This output includes the real and imaginary components of the CSI, effectively enabling improved modeling and reconstruction of wireless channel characteristics.

The following steps related to BERT4MIMO architecture are briefly explained.

\subsubsection{Input Representation and Preprocessing}

The raw CSI matrix $\mathbf{H} \in \mathbb{C}^{N_s \times N_t \times N_r}$ is split into its real and imaginary parts, and then these components are normalized independently to ensure numerical stability as mentioned in CSI matrix preprocessing.



The normalized real and imaginary parts are concatenated along a new dimension to create a unified input tensor:
\begin{equation}
    \mathbf{X} \in \mathbb{R}^{N_s \times d}, \quad d = 2 \cdot N_t \cdot N_r
\end{equation}

To handle variable-length sequences across subcarriers, zero-padding is applied to ensure uniformity across all samples. A corresponding attention mask $\mathbf{M} \in \{0,1\}^{N_s}$ is generated to denote valid and padded positions.

\subsubsection{Temporal and Feature Embedding Layers}

The input tensor $\mathbf{X}$ undergoes embedding transformations to prepare it for the transformer encoder. Each subcarrier $i$ at timestep $t$ is embedded using a combination of temporal and feature embeddings.

\paragraph{Temporal Embedding} A learnable embedding matrix $\mathbf{E}_{\text{Time}} \in \mathbb{R}^{N_s \times d_{\text{model}}}$ maps each timestep $t$ to a high-dimensional vector:
\begin{equation}
    \mathbf{e}_{t} = \text{Embedding}(t) \quad t \in \{1, 2, \ldots, N_s\}
\end{equation}

\paragraph{Feature Embedding} The feature dimension of $\mathbf{X}$ is projected into the model's latent space using a fully connected layer:
\begin{equation}
    \begin{aligned}
        \mathbf{E}_{\text{Feature}} &= \mathbf{X} \mathbf{W}_{\text{Feature}} + \mathbf{b}_{\text{Feature}}, \\
        \mathbf{W}_{\text{Feature}} &\in \mathbb{R}^{d \times d_{\text{model}}}, \quad \mathbf{b}_{\text{Feature}} \in \mathbb{R}^{d_{\text{model}}}
    \end{aligned}
\end{equation}

\paragraph{Padding and Attention Masking} Variable-length sequences in $\mathbf{X}$ are padded to a uniform length $L_{\text{max}}$. A binary attention mask $\mathbf{M}$ is created to distinguish valid tokens from padding tokens:
\begin{equation}
    \mathbf{M}[i, j] =
    \begin{cases}
        1, & \text{if token } j \text{ in sequence } i \text{ is valid}, \\
        0, & \text{otherwise}
    \end{cases}
\end{equation}

\paragraph{Combined Embedding} The temporal and feature embeddings are summed to produce the input representation for the transformer encoder:
\begin{equation}
    \mathbf{Z}_0 = \mathbf{E}_{\text{Time}} + \mathbf{E}_{\text{Feature}}, \quad \mathbf{Z}_0 \in \mathbb{R}^{N_s \times d_{\text{model}}}
\end{equation}

\subsubsection{Transformer Encoder}

The transformer encoder comprises $L$ stacked layers, each consisting of a Multi-Head Self-Attention (MHSA) mechanism and a Feedforward Network (FFN). These layers enable the model to capture global dependencies across subcarriers and antenna pairs.

\paragraph{Multi-Head Self-Attention} Each attention head computes scaled dot-product attention \cite{truong2024temporal}:
\begin{equation}
    \text{Attention}(\mathbf{Q}, \mathbf{K}, \mathbf{V}) = \text{Softmax}\left(\frac{\mathbf{Q} \mathbf{K}^\top}{\sqrt{d_k}}\right) \mathbf{V}
\end{equation}
where $\mathbf{Q}, \mathbf{K}, \mathbf{V}$ are the query, key, and value matrices, derived from the input $\mathbf{Z}_{l-1}$ at the $l$-th layer:
\begin{align}
    \mathbf{Q} = \mathbf{Z}_{l-1} \mathbf{W}_Q, \quad \mathbf{K} = \mathbf{Z}_{l-1} \mathbf{W}_K, \quad \mathbf{V} = \mathbf{Z}_{l-1} \mathbf{W}_V
\end{align}
Here, $\mathbf{W}_Q, \mathbf{W}_K, \mathbf{W}_V \in \mathbb{R}^{d_{\text{model}} \times d_k}$ are learnable parameter matrices.

The outputs from multiple attention heads are concatenated and linearly projected:
\begin{equation}
    \mathbf{H}_{\text{MHSA}} = \text{Concat}(\mathbf{H}_1, \mathbf{H}_2, \ldots, \mathbf{H}_h) \mathbf{W}_O
\end{equation}
where $h$ is the number of attention heads, and $\mathbf{W}_O \in \mathbb{R}^{(h \cdot d_k) \times d_{\text{model}}}$ is the output projection matrix.

\paragraph{Feedforward Network} The FFN applies two linear transformations with a ReLU activation \cite{pires2023one}:
\begin{equation}
    \text{FFN}(\mathbf{Z}) = \text{ReLU}(\mathbf{Z} \mathbf{W}_1 + \mathbf{b}_1) \mathbf{W}_2 + \mathbf{b}_2
\end{equation}
where $\mathbf{W}_1 \in \mathbb{R}^{d_{\text{model}} \times d_{\text{ff}}}$ and $\mathbf{W}_2 \in \mathbb{R}^{d_{\text{ff}} \times d_{\text{model}}}$

\paragraph{Residual Connections and Layer Normalization} Each sublayer output is processed with residual connections and layer normalization:
\begin{align}
    \mathbf{Z}_{l} &= \text{LayerNorm}(\mathbf{Z}_{l-1} + \text{MHSA}(\mathbf{Z}_{l-1})) \\
    \mathbf{Z}_{l} &= \text{LayerNorm}(\mathbf{Z}_{l} + \text{FFN}(\mathbf{Z}_{l}))
\end{align}

\subsubsection{Output Prediction Layer}

The final encoder output $\mathbf{Z}_L \in \mathbb{R}^{N_s \times d_{\text{model}}}$ is passed through a fully connected layer to predict the reconstructed CSI matrix:
\begin{equation}
    \hat{\mathbf{X}} = \mathbf{Z}_L \mathbf{W}_{\text{Out}} + \mathbf{b}_{\text{Out}}, \quad \mathbf{W}_{\text{Out}} \in \mathbb{R}^{d_{\text{model}} \times d}
    \; \mathbf{b}_{\text{Out}} \in \mathbb{R}^d
\end{equation}

\subsubsection{Loss Function}

The model is optimized using the Mean Squared Error (MSE) loss:
\begin{equation}
    \mathcal{L} = \frac{1}{N} \sum_{i=1}^N \|\mathbf{X}_i - \hat{\mathbf{X}}_i\|_2^2
\end{equation}
where $N$ is the number of training samples.


Algorithm~\ref{alg:BERT4MIMO_pseudo} provides a high-level description of the training process for the BERT4MIMO model.

\begin{algorithm}[ht]
\caption{BERT4MIMO Training Procedure}
\label{alg:BERT4MIMO_pseudo}
\begin{algorithmic}[1]
\REQUIRE CSI dataset $\{\mathbf{H}_i\}_{i=1}^N$, learning rate $\eta$, batch size $B$, number of epochs $E$
\STATE Initialize model parameters $\Theta$ and optimizer
\FOR{epoch $= 1$ to $E$}
    \FOR{mini-batch $\{\mathbf{H}_j\}_{j=1}^B$}
        \STATE Preprocess inputs: $\mathbf{X} \gets \text{Preprocess}(\{\mathbf{H}_j\})$
        \STATE Generate masks: $\mathbf{M} \sim \text{MaskGenerator}(B)$
        \STATE Compute predictions: $\hat{\mathbf{X}} \gets \text{BERT4MIMO}(\mathbf{X}, \mathbf{M})$
        \STATE Compute loss: $\mathcal{L} \gets \text{MSE}(\mathbf{X}, \hat{\mathbf{X}})$
        \STATE Backpropagate and update parameters: $\Theta \gets \Theta - \eta \nabla_\Theta \mathcal{L}$
    \ENDFOR
\ENDFOR
\RETURN Trained model $\Theta$
\end{algorithmic}
\end{algorithm}

\section{Experimental Results}

In this section, the performance of the proposed BERT4MIMO model is evaluated through a series of experiments, each designed to highlight different aspects of its capabilities. The experiments assess reconstruction performance, scenario-specific performance, robustness to data masking, and the impact of Doppler effects.

\subsection{Reconstruction Performance}
The reconstruction performance is used to evaluate the accuracy of a model by the reconstruction error, i.e., computing the difference between the original and reconstructed data points. The reconstruction error is typically measured using the Mean Squared Error (MSE) as follows:

\begin{equation}
\text{MSE} = \frac{1}{N} \sum_{i=1}^{N} \|\mathbf{X}[i] - \hat{\mathbf{X}}[i]\|_2^2
\end{equation}

where $N$ is the total number of samples, $\mathbf{X}[i]$ the original data, and $\hat{\mathbf{X}}[i]$ the reconstructed data. The expression $\|\|_2$ represents the Euclidean distance (L2 norm) between two vectors \(\mathbf{x}_i\) (original data point) and \(\hat{\mathbf{x}}_i\) (reconstructed data point). 

In this experiment, the accuracy of BERT4MIMO is evaluated by masking every 10th sample in the data and measuring the MSE:

\begin{equation}
\mathbf{X}_{\text{masked}}[i] = 
\begin{cases} 
\mathbf{X}[i], & \text{if } i \mod 10 \neq 0 \\
0, & \text{otherwise}.
\end{cases}
\end{equation}

According to the results, the model achieves an average MSE of $0.011035$, indicating its capability to infer missing data effectively.
However, it may not be correct to evaluate and compare the overall model performance in terms of reconstruction error due to being calculated for all scenarios and lack of similar studies at this time. 

\subsection{Scenario-wise Performance}

Scenario-wise performance is calculated with a similar way of reconstruction error using MSE for different scenarios. For each scenario, the reconstruction MSE is calculated as:

\begin{equation}
\text{MSE}_{\text{scenario}} = \frac{1}{N_{\text{scenario}}} \sum_{i=1}^{N_{\text{scenario}}} \|\mathbf{X}[i] - \hat{\mathbf{X}}[i]\|_2^2
\end{equation}

where $N_{\text{scenario}}$ is the number of samples in the scenario.

In this experiment, the performance of BERT4MIMO is evaluated across three scenarios, i.e., stationary, high-speed mobility, and urban macrocell. Table~\ref{tab:scenario_performance} presents the results, highlighting that urban macrocells pose greater challenges due to their complexity, particularly environmental such as obstacles, reflections, scattering, and noise. On the other hand, the stationary and high-speed mobility scenarios provide a better and similar performance in terms of MSE, i.e., 0.003185 and 0.003179, respectively. However, it is anticipated that the stationary scenario would provide better performance than the high-speed mobility scenario due to the operation of a stable environment.
\begin{table}[h]
    \centering
    \caption{Scenario-wise Performance of BERT4MIMO}
    \label{tab:scenario_performance}
    \begin{tabular}{|c|c|}
        \hline
        \textbf{Scenario} & \textbf{MSE} \\ \hline
        Stationary & 0.003185 \\ \hline
        High-Speed & 0.003179 \\ \hline
        Urban Macro & 0.026609 \\ \hline
    \end{tabular}
\end{table}

\subsection{Masking Ratio Sensitivity}
Masking Ratio Sensitivity is widely used in transformer-based architectures such as BERT, and typically refers to the sensitiveness of a model performance to the level of masking in input data. The masking process is defined as:

\begin{equation}
\mathbf{X}_{\text{masked}}[i] = 
\begin{cases} 
\mathbf{X}[i], & \text{if } p[i] > \gamma \\
0, & \text{otherwise}
\end{cases}
\end{equation}

where $p[i] \sim \mathcal{U}(0, 1)$ is uniformly distributed.

\begin{figure}[h]
    \centering
    \includegraphics[width=\linewidth]{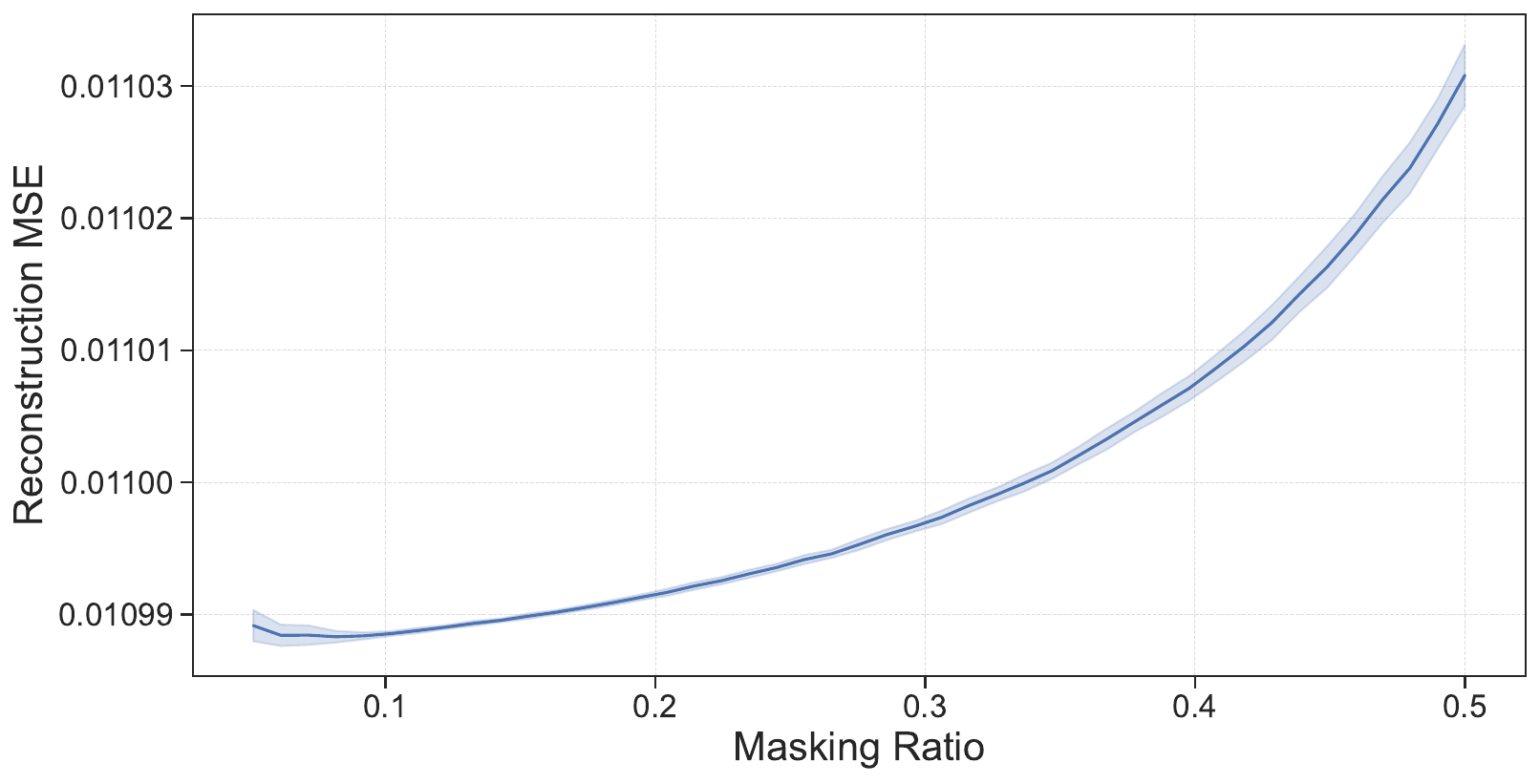}
    \caption{Effect of masking ratio on reconstruction MSE.}
    \label{fig:masking_ratio}
\end{figure}

In this study, the masking ratio $\gamma$ varies from $0$ to $0.5$ to evaluate the robustness and performance of the model at different levels of masking, i.e., missing data. Figure~\ref{fig:masking_ratio} illustrates the MSE versus $\gamma$, showing that BERT4MIMO maintains a low MSE (0.01103) for $\gamma \leq 0.5$.

\subsection{Subcarriers-wise Performance}
In this experiment, the reconstruction performance of the BERT4MIMO model is evaluated for subcarrier groups, to understand the model robustness across frequency ranges. To set up groups, the CSI data is divided into groups of 8 subcarriers. Table~\ref{tab:subcarrier_groups} presents the reconstruction performance as MSE  for each subcarrier group.

\begin{table}[h]
    \centering
    \caption{Reconstruction performance across subcarrier groups}
    \label{tab:subcarrier_groups}
    \begin{tabular}{|c|c|}
        \hline
        \textbf{Subcarrier Group} & \textbf{MSE} \\ \hline 
        0-7   & 0.012956 \\ \hline
        8-15  & 0.075252 \\ \hline
        16-23 & 0.074781 \\ \hline
        24-31 & 0.075120 \\ \hline
        32-39 & 0.076423 \\ \hline
        40-47 & 0.077504 \\ \hline
        48-55 & 0.079439 \\ \hline
        56-63 & 0.080906 \\ \hline
    \end{tabular}
\end{table}


According to the results, the subcarrier group 0-7 provides the best performance, i.e., the lowest MSE of 0.012956. It indicates that reconstruction is more accurate for these subcarriers. On the other hand,  the subcarrier group 56-63 provides the lowest performance, i.e., the highest MSE of 0.080906. Interestingly, all the subcarrier groups, except 0-7, show similar performance from 0.074781 to 0.080906, i.e., higher reconstruction error compared to the first subcarrier group (0-7). In communication systems, it means that those subcarrier groups have a high channel variability leading to the low performance. 

\subsection{Cross-Scenario Performance}
This experiment is conducted to investigate the cross-scenario performance of the model. As mentioned earlier, the dataset is divided into three scenarios as stationary, high-speed mobility, and urban macrocell. For each combination of training and testing scenarios, the model is trained exclusively with one scenario and evaluated on another. For this experiment the MSE is metric computed as folows:

\begin{equation}
\text{MSE}_{\text{cross}} = \frac{1}{N_{\text{cross}}} \sum_{i=1}^{N_{\text{cross}}} \|\mathbf{X}_{\text{test}}[i] - \hat{\mathbf{X}}_{\text{test}}[i]\|_2^2,
\end{equation}

where $N_{\text{cross}}$ is the number of test samples for each scenario.

\begin{figure}[h]
    \centering
    \includegraphics[width=1.0\linewidth]{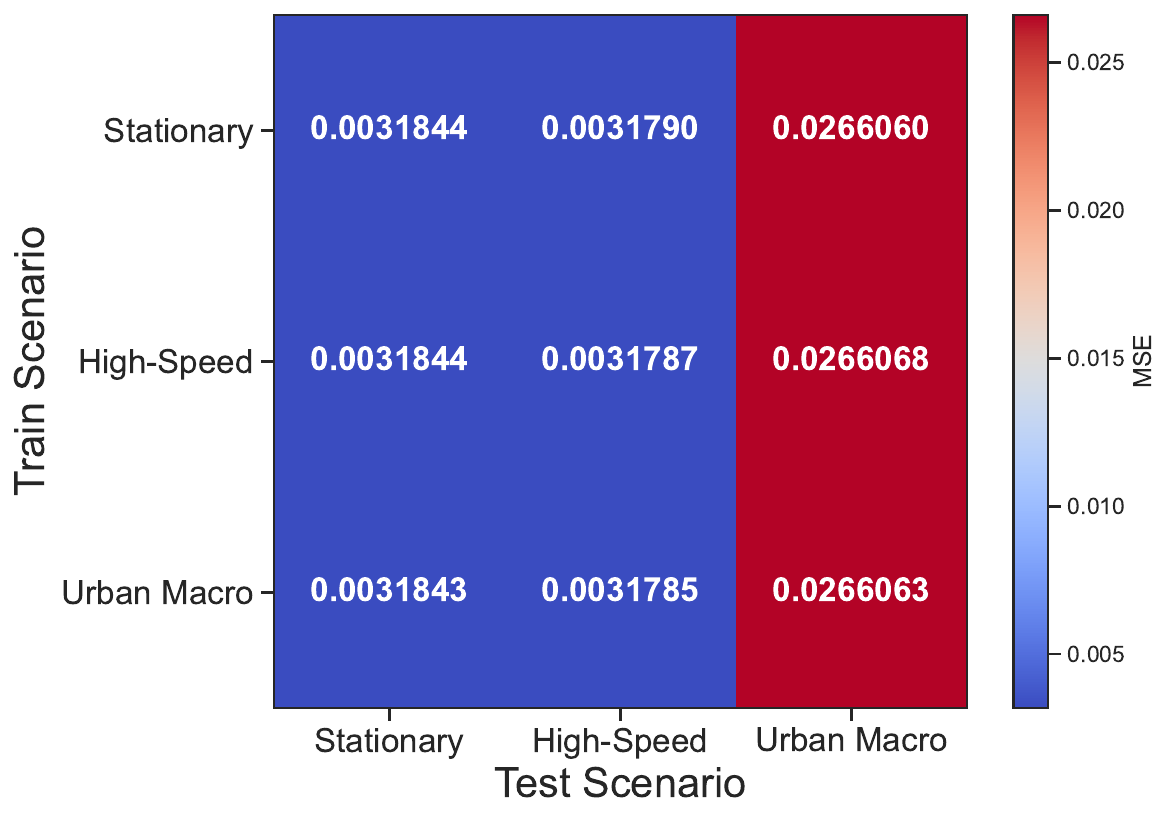}
    \caption{Heatmap of Cross-Scenario Performance. Each cell represents the MSE for a combination of training and testing scenarios.}
    \label{fig:generalization_across_scenarios}
\end{figure}

Table~\ref{tab:generalization_results} presents cross-scenario performance in terms of MSE, and the same results are visualized in Figure~\ref{fig:generalization_across_scenarios}. The diagonal entries represent the MSE when training and testing on the same scenario, while off-diagonal entries indicate cross-scenario performance.

\begin{table}[h]
    \centering
    \caption{Cross-Scenario generalization performance}
    \label{tab:generalization_results}
    \begin{tabular}{|c|c|c|c|}
        \hline
        \textbf{Train Scenario} & \textbf{Test Scenario} & \textbf{MSE} \\ \hline
        Stationary & Stationary & 0.003185 \\ \hline
        Stationary & High-Speed & 0.003182 \\ \hline
        Stationary & Urban Macro & 0.026610 \\ \hline
        High-Speed & Stationary & 0.003185 \\ \hline
        High-Speed & High-Speed & 0.003182 \\ \hline
        High-Speed & Urban Macro & 0.026611 \\ \hline
        Urban Macro & Stationary & 0.003185 \\ \hline
        Urban Macro & High-Speed & 0.003182 \\ \hline
        Urban Macro & Urban Macro & 0.026611 \\ \hline
    \end{tabular}
\end{table}

The results indicate that the BERT4MIMO model achieves consistently low MSE (i.e., a higher performance) when the model is evaluated on the same scenario, except urban macrocell scenario. In addition, the performance significantly decreases when tested on urban macrocell scenarios after being trained either on stationary or high-speed scenarios. This indicates that the complexity of urban environments intends to lower performance, and those scenarios may need fine-tuning to improve ht model performance.

\subsection{Error Distribution Analysis}
In this experiment, the model performance is investigated by analyzing the distribution of reconstruction errors across subcarrier groups. As mentioned earlier, the reconstruction error is calculated based on the difference between the original data points $\mathbf{X}$ and the constructed data points $\hat{\mathbf{X}}$ for each subcarrier group. The error distribution is analyzed for all groups using a histogram representation of errors $\epsilon$ given as follows:

\begin{equation}
\epsilon = \mathbf{X}[i] - \hat{\mathbf{X}}[i], \quad i = 1, 2, \dots, N
\end{equation}

where $N$ represents the number of samples in each subcarrier group.

\begin{figure}[h]
    \centering
    \includegraphics[width=\linewidth]{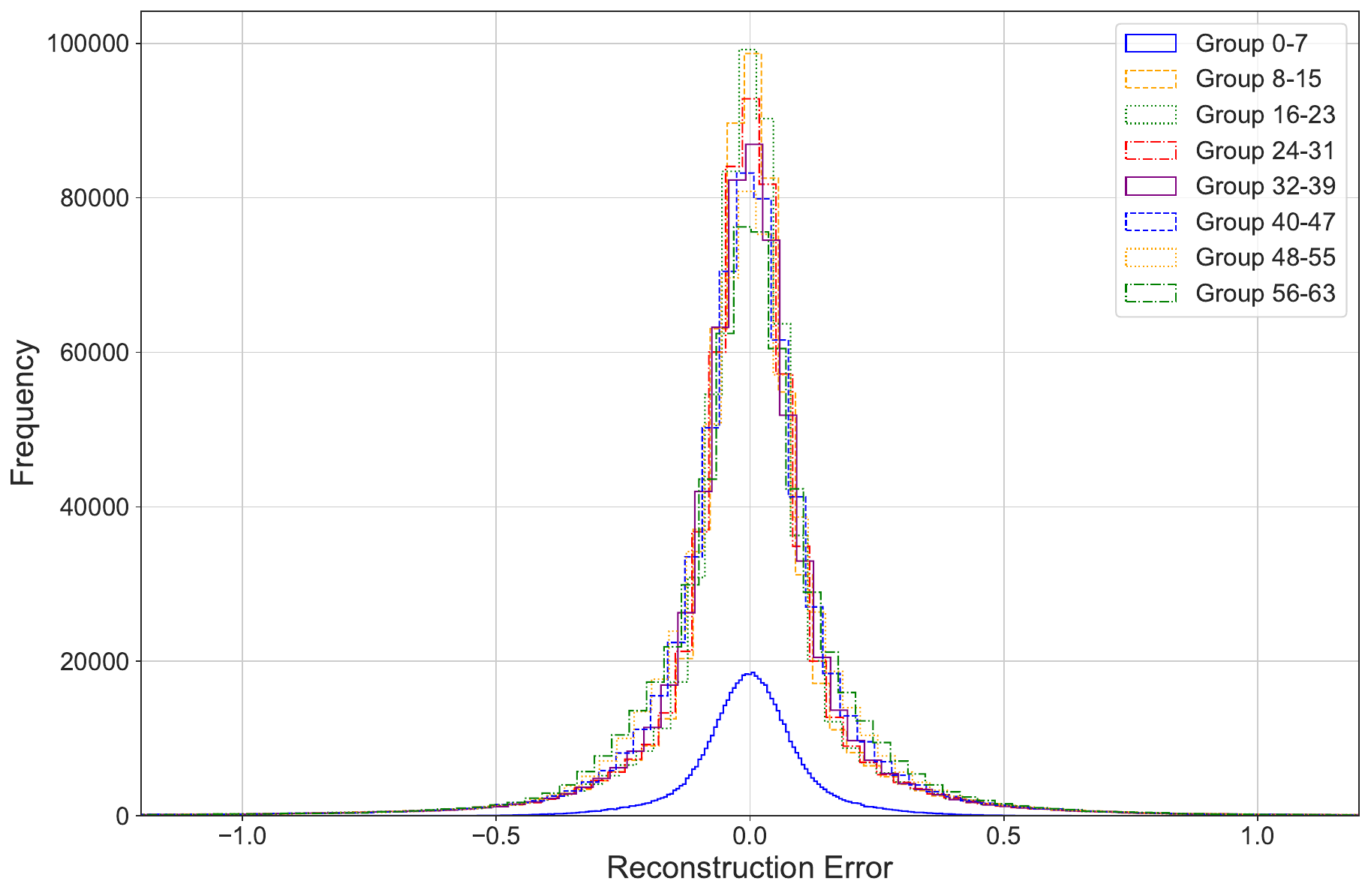}
    \caption{Error distribution across all subcarrier groups. 
    }
    \label{fig:error_distribution_all_groups}
\end{figure}

Figure~\ref{fig:error_distribution_all_groups} illustrates the distribution of reconstruction errors for each subcarrier group relative to frequency. The error distributions for individual subcarrier groups are largely uniform and show consistency across most groups. However, the distribution for subcarrier group 0-7 (blue line in the figure) is slightly narrower compared to the others, indicating lower variability in errors and highlighting BERT4MIMO's robustness. These findings emphasize the reliability of the model in a variety of channel conditions with subcarrier-specific complexities.

\subsection{Impact of Doppler Shift}
In this experiment, Doppler shifts $\delta$ are added to the CSI data based on the selected channel characteristics, and its impact on the performance of the BERT4MIMO model is evaluated. Doppler shift is one of the critical factors, particularly in scenarios with mobility due to the relative motion between the transmitter and receiver, causing frequency changes, and making the channel estimation harder.

\begin{figure}[h]
    \centering
    \includegraphics[width=0.9\linewidth]{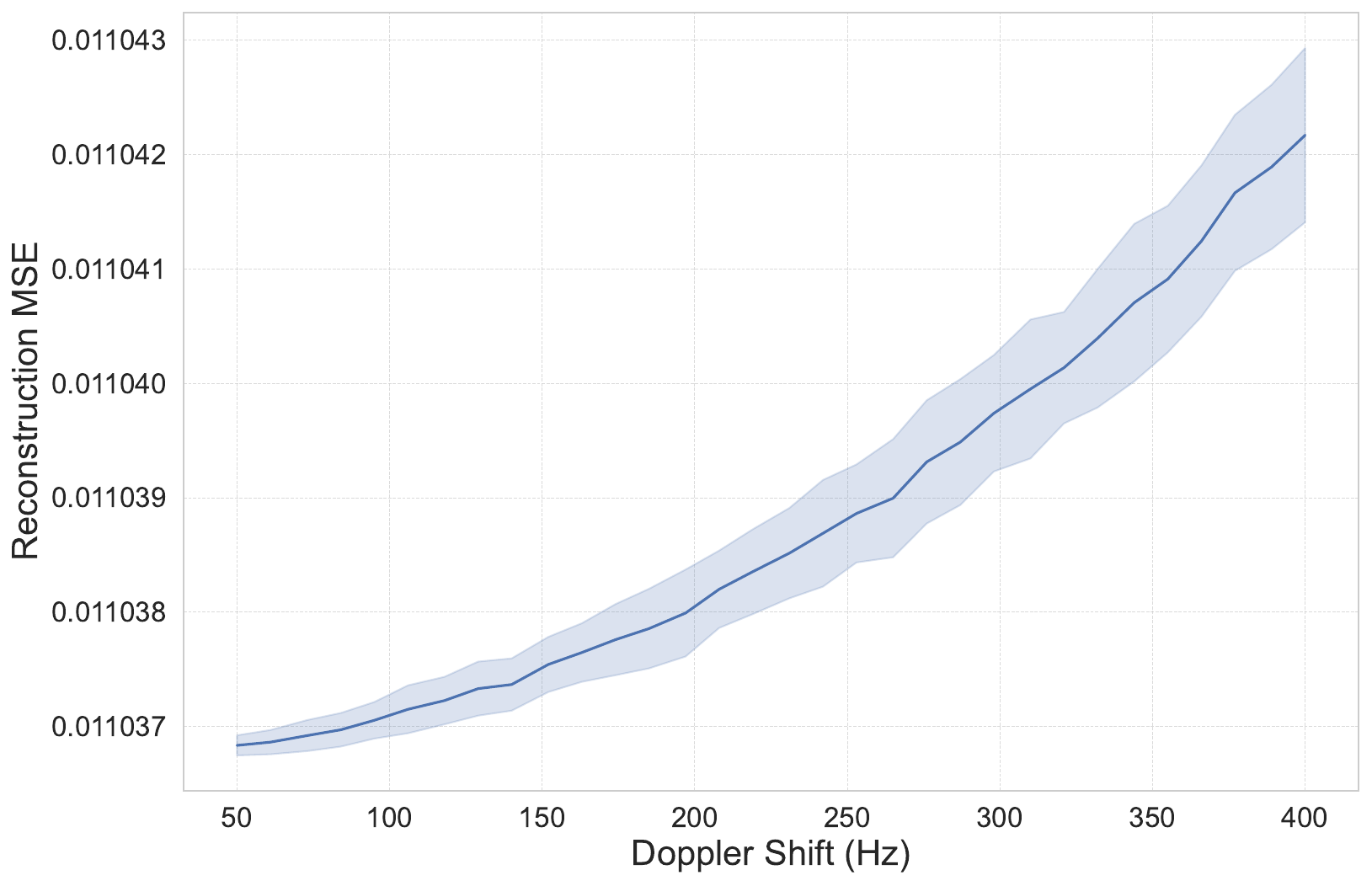}
    \caption{Impact of Doppler Shift on reconstruction MSE.}
    \label{fig:doppler_shift}
\end{figure}

Figure~\ref{fig:doppler_shift} shows the MSE as a function of $\delta$ changing from 0 Hz to 400 Hz. The results indicate that the robustness of the BERT4MIMO model reduces gradually at higher Doppler shifts, i.e., high MSE or reconstruction error. The MSE value is 0.011037 at no Doppler Shift (0 Hz) while increasing up to 0.011043 at high Doppler Shift (400 Hz).

\subsection{Baseline Comparison}
This experiment aims to compare the performance of BERT4MIMO against standard baseline models, including linear regression and a multi-layer perceptron (MLP). Two baseline models, a simple linear regression model and an MLP with a hidden layer of size 512, were trained on the masked CSI data for reconstruction. The input data $\mathbf{X}_{\text{masked}}$ and labels $\mathbf{X}$ were reshaped into two-dimensional matrices for compatibility with baseline models. The mean squared error (MSE) was computed for each model to evaluate performance.

Table~\ref{tab:baseline_comparison} summarizes the MSE values obtained for BERT4MIMO, linear regression, and MLP. It can be observed that BERT4MIMO significantly outperforms the baseline models in terms of reconstruction accuracy, achieving approximately 30 times better performance.

\begin{table}[h]
    \centering
    \caption{Baseline Comparison Results}
    \label{tab:baseline_comparison}
    \begin{tabular}{|c|c|}
        \hline
        \textbf{Model} & \textbf{MSE} \\ \hline
        BERT4MIMO & 0.011035 \\ \hline
        Linear Regression & 0.309207 \\ \hline
        MLP & 0.314465 \\ \hline
    \end{tabular}
\end{table}

The results also demonstrate the superiority of BERT4MIMO in handling the complexities of masked CSI data. While linear regression and MLP struggle to capture the intricate relationships inherent in the channel state information, BERT4MIMO leverages its attention mechanism and deep architecture to achieve a much lower reconstruction error.


\section{Observations}
The experimental results demonstrate the robustness of the BERT4MIMO as a foundation model for Massive MIMO channel state information prediction in handling complex CSI data, making it a promising approach for modern wireless communication systems. Based on the overall results, the observations can be summarized as follows:\\

\begin{itemize}
\item BERT4MIMO provides a satisfied performance with an average MSE of 0.011035 in the reconstruction of CSI data.
\item The results of the scenario-specific performance demonstrate that the urban macrocell scenario provides a lower performance (0.026609) than the stationary (0.003185) and high-speed mobility (0.003179) scenarios. This indicates that environmental complexity, such as the urban macrocell scenario, significantly impacts model performance negatively.
\item BERT4MIMO can provide a low MSE ($=<$0.01103) when the masking ratio varies from 0 to 0.5. This indicated the robustness of the model against missing data.
\item BERT4MIMO provides higher MSE or lower performance in urban macro scenarios in the cross-scenario experiment due to the environmental complexity.
\item BERT4MIMO  model provides the best MSE performance (0.012956) for the subcarrier group 0-7, while subcarrier group 56-63 has the highest MSE, i.e., 0.080906. This indicates channel variability in higher subcarriers leading to low performance.
\item Error distributions for subcarrier groups are generally uniform and consistent. However, the first subcarrier group 0-7 shows a narrower distribution, indicating lower variability at lower frequencies with reconstruction errors.
\item Doppler shift slightly increases MSE values, i.e., reducing the model performance from 0.011037 (0 Hz) to 0.011043 (400 Hz), particularly in high mobility conditions. 
\item BERT4MIMO achieves significantly higher performance (MSE: 0.011035) compared to traditional AI/ML methods such as linear regression (MSE: 0.309207) and MLP (MSE: 0.314465), i.e., approximately 30 times better. It indicates the advantages of the transformer-based architecture on a foundation model.
\end{itemize}

\section{Conclusion}
This paper presents BERT4MIMO, a novel transformer-based architecture for enhanced reconstruction of channel state information (CSI) in wireless communication systems. Utilizing the BERT model's capabilities, BERT4MIMO integrates feature embeddings to encode the temporal and spatial properties of CSI data, enabling effective modeling of complex, high-dimensional CSI data. By combining self-attention mechanisms and feedforward networks, BERT4MIMO provides a robust model for recovering and enhancing real and imaginary components of the CSI data. Comprehensive experimental evaluations for three scenarios (stationary, high-speed mobility, and urban macro cell) demonstrate the effectiveness of BERT4MIMO in reconstructing CSI data with satisfactory performance. However, the experimental results with the urban macrocell scenario usually provide a lower performance due to environmental complexity like multipath reflecting, scattering, etc. Results from the sensitivity on masking ratios indicate the model robustness against missing data. Regarding the subcarrier-wise performance, the first group (0-7) provides a better performance than other groups, indicating higher subcarriers leading to low performance due to channel variability. 


The main contribution of this study is to develop a foundation model, called BERT4MIMO, utilizing a
transformer-based architecture along with feature embeddings
to enhance CSI reconstruction for massive MIMO. For future work, the BERT4MIMO model will be extended for other applications in NextG networks, such as beamforming, spectrum sensing, IRS, AMR, and channel estimation. 

\bibliographystyle{IEEEtran}
\bibliography{refs.bib}

\end{document}